\documentclass{epl}

\usepackage{amsmath}
\usepackage{amssymb}

\title{The bend stiffness of S-DNA}
\author{C. Storm\thanks{E-mail: \email{storm@physics.upenn.edu}} \and 
P. C. Nelson}
\institute{Department of Physics and Astronomy, University of 
Pennsylvania  - Philadelphia, Pennsylvania 19104 USA
}

\def\Nxterm{\eta}          
\def\Nestiff{E}         
\def\kb{{k_{\rm B}}}
\def\kbt{{k_{\rm B}T}}

\def\tot{_{\mathrm{tot}}}
\def\eref#1{eq.~\ref{#1}}
\def\Eref#1{Eq.~\ref{#1}}

\def\pNunit{\ensuremath{\mathrm{pN}}}
\def\nmunit{\ensuremath{\mathrm{nm}}}

\def\that{\hat t}
\def\Tmat{{\sf T}}
\newcommand{\inv}{^{\raise.15ex\hbox{${\scriptscriptstyle 
-}$}\kern-.05em 1}}

\newcommand{\normsq}[1]{\|#1\|^2}

\newcommand{\cut}[1]{}
\begin{document}

\maketitle

\begin{abstract}

We formulate and solve a two-state model for the elasticity of nicked,
double-stranded DNA that borrows features from both 
the Worm Like Chain and the Bragg--Zimm model. Our model is 
computationally simple, and gives an excellent fit to recent 
experimental data through the entire 
overstretching transition. The fit 
gives the first value for the bending stiffness of the overstretched 
state as about $10\,\mathrm{nm}\cdot\kbt$, a value quite different from
either B-form or single-stranded DNA.
\end{abstract}

\section{Introduction and Summary}
When double-stranded DNA is subjected to longitudinal forces greater 
than about 65$\,\pNunit$ it undergoes a radical conformational 
change, marked by a sudden, almost twofold increase in contour 
length\cite{Cluzel,Smith}. The structural characterization of the resulting 
overstretched or (``S-state'') DNA is complicated by the fact that 
techniques such as X-ray crystallography are not applicable to single 
molecules. In this Letter, we instead characterize overstretched 
DNA by examining its elastic constants, and to this end formulate 
and solve a model that 
synthesizes features of both the  Worm Like Chain (WLC) and 
the Bragg--Zimm model of the helix--coil  transition in peptides. Thus 
we model DNA as consisting of two different, coexisting conformations, each with 
its own elastic constants. We solve this model and show that it gives 
a good fit to recent data on the overstretching 
transition in nicked, double-stranded DNA. From these fits, we 
conclude that the bend stiffness of S-DNA is intermediate between the 
known values for single stranded and double stranded DNA. Our result 
supports the work of L\'eger {\em et al.} \cite{lege99a,lege99b}, who argued 
that S-DNA has a definite helical pitch and hence is a new duplex 
conformation  of DNA.

Our model and solution method differ from those offered by Marko 
\cite{mark98a}, who assumes the bend stiffnesses of the two 
conformational states to be identical;  our analysis will show that on 
the contrary the stiffnesses are markedly different. The analysis of Viovy and Cizeau 
\cite{Acize97a} is essentially a mean-field approximation to the model 
we study here; in addition, the authors did not quote any value for the 
S-DNA bend stiffness, presumably because the  experimental data available at that 
time did not permit such a determination.

The model studied here is a continuum limit of a more general class of 
discrete persistent-chain models. Such models give better fits to the 
stretching curves of \textit{single-}stranded DNA than either the 
continuum WLC or  the freely jointed chain models. Details will appear elsewhere 
\cite{promise}.

Our model and method are also of some general interest beyond DNA. 
For example, both can be adapted to the study of the stretching of 
polypeptides with a helix-coil transition. 

\section{Model}
We begin by formulating a discretized form of the WLC, which we call 
the the ``Discrete Persistent Chain.'' Later we will introduce an 
Ising-like variable on each chain link describing a cooperative 
transition from B- to S-form.

The DPC models the polymer as a 
chain of $N$ segments
of length $b$, whose conformation is fully described by
the collection of orientation vectors $\{\hat t_i\}$ for each segment. 
Thus the  relaxed total contour length is $L_{\text{tot}}\equiv Nb$. Bend
resistance is taken into account by including an energy penalty at
each link proportional to the square of the angle
$\Theta_{i,i+1}=\arccos(\hat t_i \cdot \hat t_{i+1})$ between two 
adjacent links. The energy functional describing this model is thus given by
\begin{equation}
\frac{{\cal E}[\{\hat t_i\}]}{\kb  
T}=-\sum_{i=1}^{N}\frac{fb}{\kb
                     T} \, \hat t_i \cdot \hat z\,
+\sum_{i=1}^{N-1}\frac{A}{2b}(\Theta_{i,i+1})^2  \, .
\end{equation}
The partition function for this energy functional is  
$
{\cal Z}=\left[ \prod_{i=1}^{N} \int_{{\mathbb S}^2} \! {\rm d}^2\hat 
t_i\!\right] e^{{\cal E}[\{\hat t_i\}]/{\kb  
T}}$, where
${\mathbb S}^2$ is the two-dimensional unit sphere.

To compute ${\cal Z}$ we use the transfer matrix 
technique\cite{KramersWannier},
interpreting each integral above as a
generalized matrix product among matrices with continuous indices:
$
{\cal Z}=\vec v \cdot {\sf T}^{N-1}\cdot \vec w \, $.
In this formula $\vec v$ and $\vec w$ are vectors indexed by $\hat t$,
or in other words functions $v(\that\,),w(\that\,)$. The matrix 
product
$\Tmat\cdot\vec v$ is a new vector, defined by the convolution:
\begin{equation}
\label{dp1}
({\sf T}\cdot \vec v\,)(\hat t_i)=\int_{{\mathbb S}^2} \! {\rm 
d}^2\hat t_j
\, {\mathsf T}(\hat t_i,\hat t_j) v(\hat t_j) \, .
\end{equation}
Here the matrix elements of $\Tmat$ are given by
$
{\mathsf T}(\hat t_i,\hat t_j)=e^{-{\cal E}_i(\hat t_i,\hat t_j)/\kb  
T}\, $;
we will not need the explicit forms of $\vec v$ and $\vec w$ below.

The force-extension relation can be obtained from ${\cal Z}$ by
differentiating with respect to the force. It is here that the transfer matrix formulation can be used to greatly
simplify the calculation of the force-extension relation, since all
that is needed to compute the logarithmic derivative of ${\cal Z}$ in
the limit of long chains is the largest eigenvalue of ${\sf T}$, which
we will call $\lambda_{\text{max}}$:
\begin{equation}
\label{zldef}
\langle \frac{z}{L_{\text{tot}}} \rangle\stackrel{\mathrm{large
}N}{\longrightarrow}\left(\frac{\kb  T}{L_{\text{tot}}}\right) 
\frac{\rm
d}{{\rm d} f}
\ln (\lambda_{\max})^{N}
=\left(\frac{\kb  T}{b}\right) \frac{\rm d}{{\rm d} f} \ln 
{\lambda_{\max}}\, .
\end{equation}

Analogously to ref.~\cite{Aodij95a},
it is straightforward to add an intrinsic stretch modulus to the
calculation outlined above, obtaining an ``Extensible DPC'' model.

To study overstretching, we now extend the extensible DPC by giving 
each link a discrete variable $\sigma$, which takes
the values $\pm 1$. We will take $\sigma=+1$ to mean the segment is
in the B-state and $\sigma=-1$ for the S-state.  The factor by which 
a segment
elongates when going from B to S will be called $\zeta$, {\em
i.e.} $b^{(\mathrm{S})}=\zeta b$ (with $\zeta>1$). We assign a
bend stiffness parameter $A$ to B-DNA, and a different 
$A^{(\mathrm{S})}\equiv\beta
\zeta A$ to S-DNA; $\beta$ is a dimensionless parameter with 
$\beta\zeta<1$.
Similarly we assign a bend stiffness $\Nxterm A$ to a
hinge joining a B and an S segment. 

The full energy functional for the Ising--DPC model is thus:
\begin{eqnarray}
\label{idpcen}
\frac{{\cal E}[\{\hat t_i,\sigma_i\}]}{\kb  T}&=&-\sum_{i=1}^{N-1} 
\biggl\{
\frac{\alpha_0}{2}(\sigma_i\!+\!\sigma_{i+\!1})+\gamma(\sigma_i
\sigma_{i+\!1}\!-\!1)\nonumber+
\\&&\hspace{-1.5cm}+\frac{fb}{2\kb
T}\left[\Bigl(\frac{1\!+\!\sigma_i}{2}\!+\!\frac{1\!-\!\sigma_i}{2}\zeta\Bigr)\hat t_i \cdot 
\hat z\!+\!\Bigl(\frac{1\!+\!\sigma_{i+\!1}}{2}\!+\!\frac{1\!-\!\sigma_{i+\!1}}{2}\zeta\Bigr)
\hat t_{i+\!1}\cdot \hat z\right]\!- \nonumber \\
& &\hspace{-.6cm} -\frac{A}{2b}\left[
\frac{(1\!-\!\sigma_i)(1\!-\!\sigma_{i+\!1})}{4}\beta+|\sigma_i\!-\!\sigma_{i+\!
1}|\Nxterm+
\frac{(1\!+\!\sigma_i)(1\!+\!\sigma_{i+\!1})}{4}\right]
(\Theta_{i,i+\!1})^2 \biggr\} \, .
\end{eqnarray}
The first line is the pure-Ising part, with $2 \alpha_0 \kb T$ the
intrinsic free energy cost of converting a single segment from B to S 
and
$2\gamma \kb T$ the energy cost of creating
a B$\to$S interface. Note that we ignore a contribution to the energy
functional from the first and last segments. In the long-chain limit 
this
does not affect the outcome of our calculation.

The partition function for the energy functional (\ref{idpcen}) 
is given by
\begin{equation}
\label{2dpcZ}
{\cal Z}=\left[\prod_{i=1}^{N-1}\sum_{\sigma_i=\pm 1} \! 
\int_{{\mathbb
S}^2} \! {\rm d}^2\hat t_i\right] \!  \prod_{i=1}^{N-1}e^{-{\cal 
E}_i(\hat
t_i,\sigma_i,\hat t_{i+1},\sigma_{i+1})/\kb  T}\, ,
 \end{equation}
where now ${\cal
E}[\{ \hat t_i,\sigma_i \}]=\sum_{i=1}^{N-1}{\cal E}_i(\hat
t_i,\sigma_i,\hat t_{i+1},\sigma_{i+1})$. 
We again calculate ${\cal Z}$ with the aid of the transfer matrix
technique, writing \eref{2dpcZ} as
$
{\cal Z}=\vec v \cdot {\sf T}^{N-1}\cdot \vec w\, $,
with ${\sf T}$ now the transfer matrix for our Ising-DPC model,
which carries an additional 2-by-2 structure due to the Ising
variables. The dot products are thus defined as
\begin{equation}
({\sf T}\cdot \vec v)_{\sigma_i}(\hat t_i)=\sum_{\sigma_j=\pm 1} \!
\int_{{\mathbb S}^2} \! {\rm d}^2\hat t_j \, {\mathsf T}_{\sigma_i
\sigma_j}(\hat t_i,\hat t_j) v_{\sigma_j}(\hat t_j) \, .
\end{equation}
The individual matrix elements ${\mathsf T}_{\sigma_i \sigma_j}$ are 
given explicitly by
\begin{eqnarray*}\label{Telts}
{\mathsf T}_{-1,-1}(\hat t_i,\hat t_{i+1})& = & \exp\left[\frac{1}{2} 
\zeta
\tilde f(\hat
t_i\!+\!\hat t_{i+1})\cdot \hat z\!-\!\frac{\beta A}{b}(1\!-\!\hat t_i
\cdot \hat
t_{i+1})\!-\!\alpha_0\right]\, ,
\end{eqnarray*}
and related expressions for ${\mathsf T}_{1,1}$, ${\mathsf T}_{1,-1}$, and ${\mathsf T}_{-1,1}$,
where $\tilde f\equiv\frac{fb}{\kb  T}$.

We approximate the largest eigenvalue of the
transfer matrix ${\sf T}$  using a variational approach\cite{MarkoSiggia}. 
We choose a three-parameter family of trial eigenfunctions with azimuthal symmetry,
peaked in the direction of the force $\hat z$:
\begin{equation} \label{eomphi}
 v_{\omega_{1},\omega_{-1},\varphi}(\hat t\,)=
\left(\begin{array}{c } \left(\frac{\omega_{1}}{\sinh(2 
\omega_1)}\right)^{\frac12} e^{\omega_1 \hat t \cdot \hat z} \cos 
\varphi \\
 \left(\frac{\omega_{-1}}{\sinh(2 
\omega_{-1})}\right)^{\frac12}e^{\omega_{-1} \hat t \cdot \hat z}\sin 
\varphi \end{array}\right)\, .
\end{equation}
These trial functions were
chosen such that their squared norm is independent of all parameters:
$
\| \vec v_{\omega_{1},\omega_{-1},\varphi}\|^2=2 \pi$.
\Eref{eomphi} shows that the $\omega$'s give the degree of
alignment of the monomers (how forward-peaked their probability
distribution is), whereas $\varphi$ describes the relative probability
of a monomer to be in the two states. The variational estimate for the
maximal eigenvalue is thus
\begin{equation}\label{varesti}
\lambda^*_{\text{max}}\equiv\max_{\omega_{1},\omega_{-1},\varphi} 
y(\omega_{1},\omega_{-1},\varphi)
\equiv\max_{\omega_{1},\omega_{-1},\varphi}\, \frac{\vec 
v_{\omega_{1},\omega_{-1},\varphi} \cdot
{\sf T} \cdot \vec v_{\omega_{1},\omega_{-1},\varphi}
}{\normsq{\vec v_{\omega_{1},\omega_{-1},\varphi}}} \, ,
\end{equation}

The maximization over $\varphi$ can be done analytically: defining the
$2\times 2$ matrix $\tilde {\sf T}(\omega_1,\omega_{-1})$ by
\begin{equation}
\vec v_{\omega_{1},\omega_{-1},\varphi} \cdot {\sf T} \cdot \vec
v_{\omega_{1},\omega_{-1},\varphi}=(\cos \varphi,\sin \varphi) \cdot 
\tilde {\sf
T}(\omega_1,\omega_{-1}) \cdot \left(\begin{array}{c }   \cos \varphi 
\\ \,
\sin \varphi \end{array}\right) \, ,
\end{equation}
gives that
\begin{equation}\label{e:lstmax}
\lambda^*_{\text{max}}=\max_{\omega_1,\omega_{-1}}\, \frac{\tilde 
y(\omega_1,\omega_{-1})}{\normsq{\vec
v_{\omega_{1},\omega_{-1},\varphi}}}\, ,
\end{equation}
where $\tilde y(\omega_1,\omega_{-1})$ is the maximal eigenvalue of 
the $2\times2$ matrix
$\tilde{\sf T}(\omega_1,\omega_{-1})$. The following section will 
calculate this
eigenvalue in a
continuum approximation to $\tilde{\sf T}(\omega_1,\omega_{-1})$,  
illustrating the
procedure by considering in some detail the matrix element 
$\tilde{\mathsf T}_{1,1}(\omega_1,\omega_{-1})$.  The other matrix elements can be 
obtained analogously.
Writing out the integrals explicitly, we have
\begin{equation}
\tilde{\mathsf T}_{1,1}(\omega_1)=\frac{\omega_{1} 
e^{\alpha_0-\frac{A}{b}}}{\sinh(2 \omega_1)}\int_{{\mathbb
S}^2} \! {\rm
d}^2\hat t_i e^{\hat a \hat t_i \cdot \hat z}\int_{{\mathbb S}^2} \! 
{\rm
d}^2\hat t_{i+1}
\left[e^{(\hat a \hat z+\frac{A}{b}\hat t_i)\cdot \hat 
t_{i+1}}\right]\, ,
\end{equation}
where we have introduced $\hat a \equiv \omega_1+\frac{\tilde f}{2}$.
Condensing notation even further we define $\mu^2=\hat
a^2+(\tfrac{A}{b})^2+2\hat a\tfrac{A}{b}\hat t_i \cdot \hat z$, which
allows us to write
\begin{equation}\label{eqT}
\tilde{\mathsf T}_{1,1}(\omega_1)\!=\!(2
\pi)^2\frac{\omega_{1} e^{\alpha_0-{A}/{b}}}{\sinh(2 
\omega_1)}\int_{|\tfrac{A}{b}\!-\!\hat a|}^{\tfrac{A}{b}\!+\hat 
a}\!\!  \
\frac{ b\,{\rm d}\mu}{\hat a A} e^{{b}(\mu^2\!-\!\hat
a^2\!-\!(\frac{A}{b})^2)/{(2 A)}}\left[ e^\mu\!-\!e^{-\mu}\right]\, .
\end{equation}

\section{Continuum Limit\label{s:cl}}
We could now proceed to evaluate the force-extension relation of the
Ising-DPC model, by evaluating \eref{e:lstmax} numerically and using \eref{zldef}. To simplify the
calculations, however, we first pass to a continuum limit. To
justify this step, we note that the
continuum (WLC) approximation gives a good account of
single-stranded DNA stretching out to forces beyond those probed
in overstretching experiments (about $90\,\pNunit$) \cite{Clausen}.  As mentioned
earlier, the continuum approximation is also quite good for 
double-stranded
DNA, because the latter's persistence length is so much longer than
its monomer size.

In the continuum limit $b$ is sent to zero holding $L\tot$ fixed; 
hence
$N\to\infty$. The bookkeeping is more manageable after a shift in
$\mu$:
$
x\equiv \mu-({A}/{b})$.
\Eref{eqT} then reduces to
\begin{eqnarray}
\tilde{\mathsf T}_{1,1}(\omega_1)\!&=&\!\frac{\omega_{1} 
e^{\alpha_0}}{\sinh(2 \omega_1)} \frac{(2 \pi)^2 b}{\hat a
A}\int_{-\hat a}^{+\hat a} \!\!  \, {\rm d}x\, \exp \left[ \frac{b}{2 
A}
x^2 +2x-\frac{\hat a^2 b}{2A}\right] \nonumber \\
&\approx& \!\frac{\omega_{1} e^{\alpha_0}}{\sinh(2 \omega_1)} 
\frac{(2 \pi)^2
b}{\hat a A}\int_{-\hat a}^{+\hat a} \!\!  \, {\rm d}x\, 
e^{2x}(1+\frac{x^2
b}{2 A}) e^{-\frac{\hat a^2 b}{2 A}}\, .
\end{eqnarray}
The last integral can be worked out exactly, and expanding the result 
to second order in $b$ we end up with 
\begin{equation}
\frac{A}{2\pi b} \frac{1}{\normsq{\vec 
v_{\omega_{1},\omega_{-1},\varphi}}}\tilde{\mathsf T}_{1,1}(\omega_1)\!=\!e^{\alpha_0}\left[ 1+{b}\left(\frac{f}{\kb 
T}-\frac{\omega_1}{2A}\right)\left(\coth(2\omega_1)-\frac{1}{2\omega_1}\right)\right]\, 
.
\end{equation}
In similar fashion, we can obtain the following expressions for the 
other
matrix elements.
\begin{eqnarray}
\frac{A}{2\pi b} \frac{1}{\|\vec 
v_{\omega_{1},\omega_{-1},\varphi}\|^2}\tilde{\mathsf T}_{-1,-1}(\omega_{-1})\!&=&\!\beta\inv e^{-\alpha_0}\left[
1+{b}\left(\frac{\zeta
f}{\kb  T}-\frac{\omega_{-1}}{2 \beta
A}\right)\left(\coth(2\omega_{-1})-\frac{1}{2\omega_{-1}}\right)\right] 
\nonumber \\
\frac{A}{2\pi b} \frac{1}{\|\vec 
v_{\omega_{1},\omega_{-1},\varphi}\|^2}\tilde{\mathsf T}_{1,-1}(\omega_1,\omega_{-1})\!&=&\!\frac{e^{-2\gamma}}{\Nxterm} 
\left( \frac{\omega_1 
\omega_{-1}}{\sinh(2\omega_1)\sinh(2\omega_{-1})}\right)^{\frac12} 
\left( \frac{2 
\sinh(\omega_1+\omega_{-1})}{\omega_1+\omega_{-1}}\right)\,.
\end{eqnarray}

To obtain a nontrivial continuum limit we must now specify how the
parameters $A$, $\alpha_0$, and $\gamma$ depend on $b$ as $b \to 0$. 
The choices
\begin{equation}
\alpha_0=-\frac12\ln\beta+b\bar\alpha\,,\qquad \gamma=-\frac12\ln(\bar
g b)
\end{equation}
give a well-defined limit, where we hold $A$, $\bar\alpha$, $\beta$ and $\bar g$ fixed as 
$b\to0$.
With these choices, the matrix
$\frac1{\normsq{\vec v_{\omega_{1},\omega_{-1},\varphi}}}\tilde{\sf 
T}(\omega_{1},\omega_{-1})$ takes the form
\begin{equation}
\label{tmatcons}
 \frac{1}{\|\vec v_{\omega_{1},\omega_{-1},\varphi}\|^2}\tilde{\sf 
T}(\omega_{1},\omega_{-1})=\frac{2 \pi
b}{A \sqrt{\beta}}\left({\mathsf 1} +b
\begin{pmatrix}
    {\cal P}  & {\cal Q}    \\
 {\cal Q}     & {\cal R}
\end{pmatrix}\right)\, ,
\end{equation}
with
\begin{eqnarray}
{\cal P}&=&\bar\alpha  +
\left(\frac{f}{\kb
T}-\frac{\omega_1}{2A}\right)\left(\coth(2\omega_1)-\frac{1}{2\omega_1}\right)\nonumber\,
,\\
{\cal R}&=&-\bar\alpha  + \left(\frac{\zeta
f}{\kb T}-\frac{\omega_{-1}}{2A\beta
}\right)\left(\coth(2\omega_{-1})-\frac{1}{2\omega_{-1}}\right)\nonumber 
\, ,\\
{\cal Q}&=&\frac{\bar g \sqrt{\beta}}{\Nxterm}\left( \frac{\omega_1 
\omega_{-1}}{\sinh(2\omega_1)\sinh(2\omega_{-1})}\right)^{\frac12} 
\left( \frac{2 
\sinh(\omega_1+\omega_{-1})}{\omega_1+\omega_{-1}}\right)\, .
\end{eqnarray}
Note that the prefactor $\frac{2\pi b}{A \sqrt\beta}$ in 
\eref{tmatcons} does not
contribute to the force-extension result \eref{zldef}, because it
does not depend on the force.  In terms of the individual matrix
entries, the quantity to be maximized now reads (see \eref{varesti}):
\begin{equation}
\label{lamom}
\ln \tilde y(\omega_{1},\omega_{-1})=\frac{b}{2}\left({\cal P}+{\cal
R}+\sqrt{({\cal P}-{\cal R})^2+4{\cal Q}^2}\right)\, .
\end{equation}
Writing $\Omega\equiv b\inv\ln\lambda^*_{\text{max}}=
b\inv\times\max \ln
\tilde y(\omega_{1},\omega_{-1})$, the force-extension in the 
continuum limit is finally
given by
\begin{equation}
\langle \frac{z}{L_{\rm tot,b}}\rangle=\kb  T\frac{{\rm 
d}\Omega}{{\rm d}
f} \, .
\end{equation}
We evaluate $\Omega$ by numerically maximizing \eref{lamom}.

\begin{figure}[b!]
\begin{center}
  \includegraphics[width=4truein]{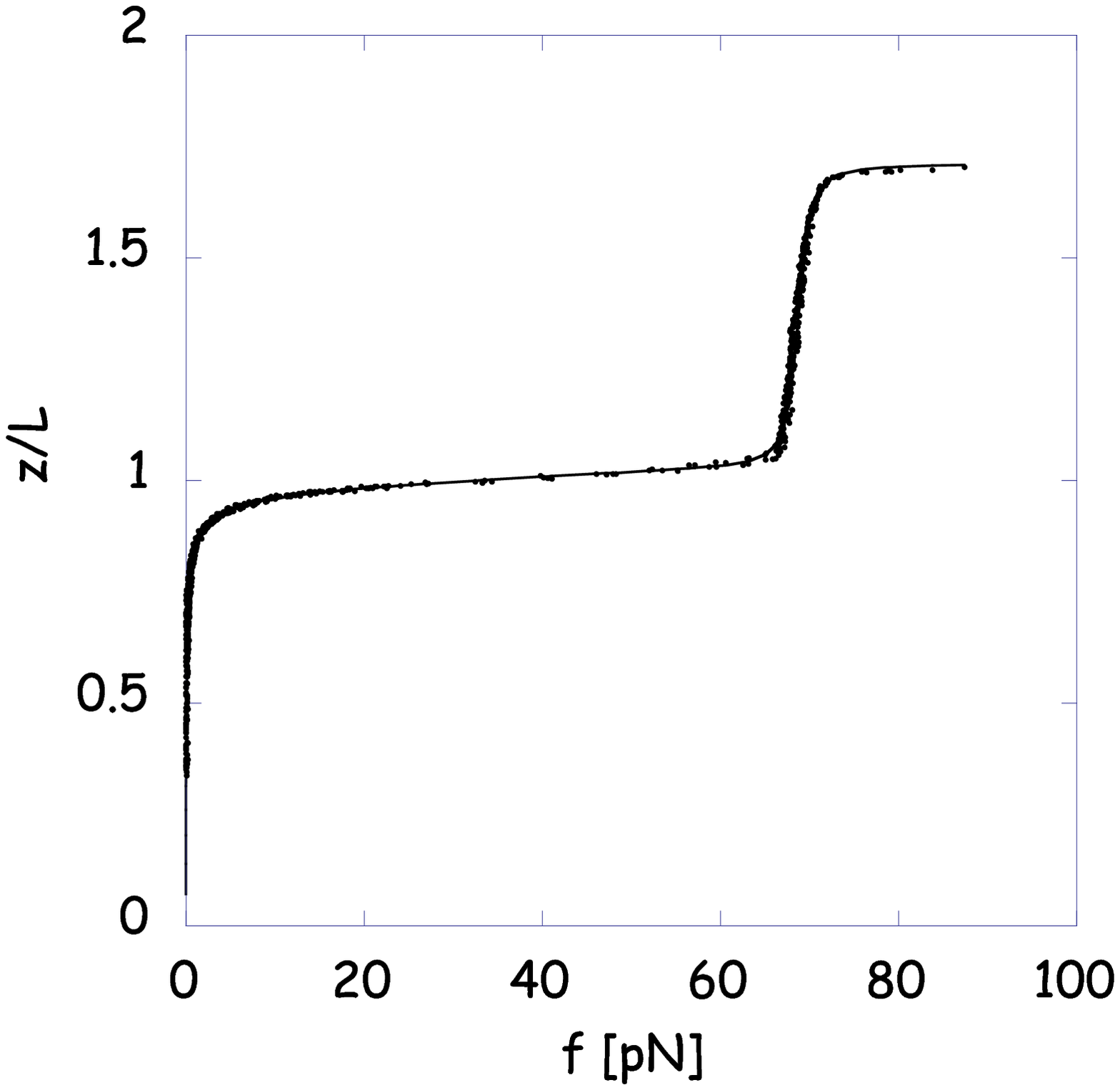}
\end{center}
\caption[]{Least-squares fit of the Ising-DPC model to an overstretching
dataset  (48.5\thinspace kbp $\lambda$
DNA construct; buffer 500\thinspace mM NaCl,
20\thinspace mM Tris, pH~8). Data kindly supplied by C.~Bustamante and
S.~Smith.
Fit parameters: $A=43.75\,\nmunit$,
$\bar\alpha =5.45\, \nmunit^{-1}$, $\beta=0.16$, ${\cal 
Q}=0.13\, \nmunit^{-1}$, $\zeta=1.76$,
$\Nestiff^{(\mathrm B)}=1.2\cdot 10^3\,\pNunit$ and
$\Nestiff^{(\mathrm S)}=1.0\cdot10^4\,\pNunit$. $\chi^2=9.22$ at $N=825$; points 
with $1.11<\langle\frac{z}{L}\rangle < 1.55$ were excluded from the fit.\label{smithfit}}
\end{figure}

So far, we have not included stretch moduli for the B- and
S-DNA. This is easily implemented to first order in $f/\Nestiff$ by
replacing $f$
with $f(1+\frac{f}{2 \Nestiff^{(\mathrm{S,B})}})$ in the matrix elements for the 
two
states respectively (\eref{Telts}). This procedure yields theoretical
force-extension curves like the one plotted in 
Fig.~(\ref{smithfit}).

In summary, our model contains the following seven parameters.  $2
\bar \alpha \kb T$ is the free energy per unit length required to 
flip B-DNA into
the S-state, and is measured in [J/nm].  ${\cal Q}$ measures
the cooperativity of the transition and has units [1/nm].  $A$ is the
bend stiffness parameter of B-DNA, with units [nm].  The 
dimensionless parameter $\beta$
is the ratio of the B- and S-DNA bend stiffnesses.  $\Nestiff^{(\mathrm{B})}$ and
$\Nestiff^{(\mathrm{S})}$ are the stretch stiffnesses of B and S-DNA, and are
measured in pN. Finally, $\zeta$ is the dimensionless elongation
factor associated with the B$\to$S transition.
\section{Discussion of fit\label{fitdisc}}
Fig.~(\ref{smithfit}) shows a fit to some recent experimental data 
(similar data appear in \cite{bust01a}).  Our 
model reproduces the experimental data rather well, but with so many fit parameters one
may ask whether it actually makes any falsifiable predictions.
To answer this question we note that the data below the transition
suffice to fix $A$ and $\Nestiff^{(\mathrm{B})}$ as usual, roughly speaking from 
the
curvature and slope of the curve below the transition. Similarly,
the data above the transition fix $A^{(\mathrm{S})}=\zeta\beta A$ and
$\Nestiff^{(\mathrm{S})}$. The vertical jump in the curve at the transition fixes
$\zeta$. The horizontal location of the jump fixes $\bar\alpha$, and
the steepness of the jump fixes the cooperativity ${\cal
Q}$.\footnote{The fit value of $\bar\alpha$ should be regarded as an
average of the two different costs to convert AT or GC pairs. The fit
value of ${\cal Q}$ has no direct microscopic significance, as the
apparent cooperativity of the transition will be reduced by the
sequence disorder.} Thus all
of the model's parameters are fixed by specific features of the data.
Two additional, independent features of the data now remain, namely 
the
rounding of the curve at the start and end of the transition. Our
model predicts these features fairly succesfully.

The fit recovers the
known values for the effective persistence length of B-DNA of around
$50\,$nm and its stretch modulus of about $1000\,$pN. Our first result is that the bend stiffness of S-DNA from our fit as
$A^{(\mathrm{S})}=\beta\zeta A=12.32\,$nm. Similar results were 
obtained using the older data of Cluzel {\it et al.} \cite{Cluzel,promise}.
If S-DNA consisted of two unbound, single
strands, we might have expected $A^{(\mathrm{S})}$ to be twice as large as the 
value
$A^{\rm ss}\approx0.75\,$nm appropriate to single-stranded DNA (as obtained from stretching experiments on ssDNA, restricted to forces above those required to pull out secondary structure \footnote{Hagerman's result that the persistence length of a single strand of poly(dT) DNA is between 2 nm and 3 nm \cite{Hagerman} does not come from a stretching experiment and should not be compared directly.}  \cite{Smith,Clausen,promise}). On the contrary, we 
find that the bend stiffness of S-DNA is intermediate between 
that of
B-DNA and that of
two single strands. 

Finally, our fit gives the stretch modulus of S-DNA is
substantially higher than that of B-DNA. This conclusion is  
consistent with the idea that the contour length of S-DNA is 
determined by its covalently bonded sugar-phosphate
backbones, which are much straighter than in B-DNA; the contour length
of B-DNA is instead determined by weaker, base-stacking interactions.

\section{Relation to prior work}
Several authors have studied the entropic elasticity of
two-state chains. As soon as the overstretching transition was
discovered, Cluzel proposed a pure Ising model by analogy to the
helix-coil transition \cite{cluz96a}. Others then introduced entropic
elasticity, but required that both states have the same bending
stiffness as B-DNA \cite{mark98a,Ahsan} or took one of the
two states to be
infinitely stiff \cite{tama01}, or to be a FJC
\cite{RouzinaBloomfield1}. Also several earlier works made a 
mean-field approximation instead of diagonalizing the full transfer 
matrix. We believe our Ising-DPC
model to be the first consistent formulation incorporating the
coexistence of two different  states with arbitrary elastic constants.
Our approach also is calculationally  more straightforward than some,
and minimal in the sense that no unknown potential function needs to
be chosen.

\acknowledgments
We thank T. Burkhardt, D. Chatenay, A. Grosberg, R. Kamien, J. Marko 
and M. Rief for valuable
discussions, and  C.~Bustamante, D.~Chatenay, J.-F. L\'eger, J. 
Marko, M.
Rief, and S. Smith for sending us experimental data.
CS acknowledges support from NIH grant R01 HL67286 and from NSF grant 
DMR00-79909. 
PN acknowledges partial support from NSF grant DMR98-07156.


\end{document}